\documentclass[journal,romanappendices]{IEEEtran}
\usepackage{amssymb}
\usepackage[cmex10]{amsmath}
\usepackage{overpic}
\usepackage{algorithm}
\usepackage{algorithmic}
\usepackage{multirow}
\usepackage{array}
\usepackage{graphicx}
\usepackage{subfigure}
\usepackage{bm}
\usepackage[caption=false,font=normalsize,labelfont=sf,textfont=sf]{subfig}
\usepackage{color}
\usepackage{tikz}
\usepackage{epstopdf}
\usepackage{stfloats}
\usepackage{cite}
\usepackage{tikz}
\usepackage{amsmath}
\usepackage{psfrag}
\usepackage{acronym}
\usepackage{enumerate}
\usetikzlibrary{arrows}
\usetikzlibrary{decorations}

\definecolor{BLUE}{rgb}{0,0,1}

\acrodef{crb}[CRB]{Cram\'{e}r-Rao bound}
\acrodef{mse}[MSE]{mean-squared error}
\acrodef{nisq}[NISQ]{noisy intermediate-scale quantum}
\acrodef{rmse}[RMSE]{root-mean-squared error}
\acrodef{qem}[QEM]{quantum error mitigation}
\acrodef{qpr}[QPR]{quasi-probability representation}
\acrodef{kkt}[KKT]{Karush-Kuhn-Tucker}

\usepackage[braket]{qcircuit}
\usepackage{winsnotation}
\usepackage{mathrsfs}
\usepackage{rotating}
\newcommand{\Sop}[1]{{\mathcal{#1}}}

\begin{document}
\title{\huge Dual-Frequency Quantum Phase Estimation Mitigates the Spectral Leakage of Quantum Algorithms}

\author{Yifeng Xiong, \IEEEmembership{Student Member, IEEE}, Soon Xin Ng, \IEEEmembership{Senior Member, IEEE}, Gui-Lu Long, \IEEEmembership{Member, IEEE}, \\ and Lajos Hanzo, \IEEEmembership{Fellow, IEEE}
\thanks{The authors are with School of Electronics and Computer Science, University of Southampton, SO17 1BJ, Southampton (UK).}
\thanks{L. Hanzo would like to acknowledge the financial support of the Engineering and Physical Sciences Research Council projects EP/P034284/1 and EP/P003990/1 (COALESCE) as well as of the European Research Council's Advanced Fellow Grant QuantCom (Grant No. 789028). This work is also supported in part by China Scholarship Council (CSC).}}

\maketitle

\begin{abstract}
Quantum phase estimation is an important component in diverse quantum algorithms. However, it suffers from spectral leakage, when the reciprocal of the record length is not an integer multiple of the unknown phase, which incurs an accuracy degradation. For the existing single-sample estimation scheme, window-based methods have been proposed for spectral leakage mitigation. As a further advance, we propose a dual-frequency estimator, which asymptotically approaches the Cram\'{e}r-Rao bound, when multiple samples are available. Numerical results show that the proposed estimator outperforms the existing window-based methods, when the number of samples is sufficiently high.
\end{abstract}


\section{Introduction}
\IEEEPARstart{Q}{uantum phase estimation} \cite{qpe,qpe2,qpe3} is a key enabler of quantum computational speedup over classical computers. It is a widely used component of quantum algorithms providing substantial acceleration over their best classical counterpart, including Shor's factoring algorithm \cite{shor}, the quantum clock synchronization algorithm \cite{qpe_sync,qpe_sync_exp}, the Harrow-Hassidim-Lloyd (HHL) algorithm \cite{hhl,vtc_hhl} conceived for solving linear systems, and the quantum counting algorithm \cite{qcounting,qaa}. Recently, iterative quantum phase estimators have also been proposed for potential application in near-term \ac{nisq} computers \cite{iqpe,iqpe2,iqpe3}.

\begin{figure}[t]
\centering
\includegraphics[width=.485\textwidth]{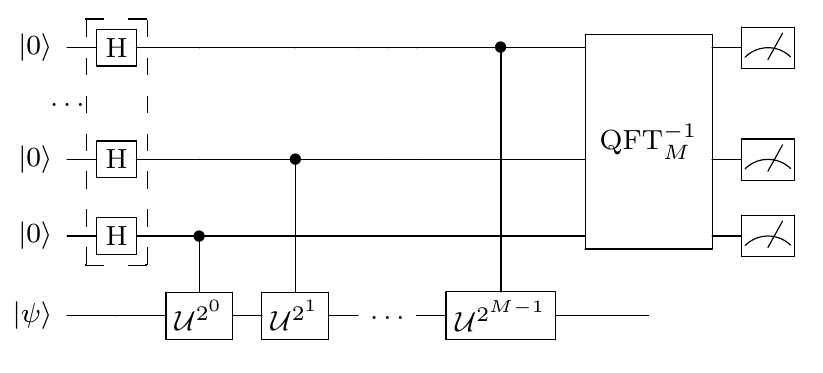}
\caption{The circuit diagram of quantum phase estimation.}
\label{fig:qpe_diagram}
\end{figure}

The circuit diagram of quantum phase estimation is portrayed in Fig.~\ref{fig:qpe_diagram}, where the initial state $\ket{\psi}$ of the data register is an eigenstate of the unitary operator $\Sop{U}$ satisfying $\Sop{U}\ket{\psi} = e^{j\varphi}\ket{\psi}$, and $\mathrm{QFT}_M^{-1}$ denotes the inverse quantum Fourier transform \cite[Sec. 5.1]{ncbook} applied to $M$ qubits, which is a quantum-domain version of the classical discrete Fourier transform \cite[Sec. 8]{dsp}. This algorithm aims for estimating the phase $\varphi$. Broadly speaking, the $2^M-1$ controlled unitaries produce a $2^M$-point sinusoidal signal in the control register, whose frequency corresponds exactly to the desired phase $\varphi$. This may be viewed as the quantum-domain version of the power method of eigendecomposition \cite{GoluVanl96}. When efficient implementations of the power of the unitary operator $\Sop{U}$ are available, exponential speedup over classical algorithms is possible, as seen in Shor's algorithm \cite{shor}.

Quantum phase estimation yields the exact result, when the recording length $2^M$ is an integer multiple of the period of the sinusoidal signal \cite[Sec. 5.2.1]{ncbook}. When this is not the case, the spectral leakage problem \cite{spectral_leakage} arises, which is an important topic in classical signal processing methods related to the discrete Fourier transform. The quantum-domain version of this problem is somewhat more grave, since the measurement outcome would assume erroneous values at non-negligible probabilities, hence the information about the correct phase is lost.

One of the conventional solutions to the spectral leakage issue is to multiply the time-domain signal by a smooth ``window function'' \cite{spectral_leakage}. This idea has been applied to the quantum phase estimation problem in \cite{quantum_cosine}, where the authors show that the cosine window \cite[Sec. 7]{dsp} is optimal in terms of the \ac{mse} of single-sample-based estimation. An efficient quantum circuit-based implementation of the cosine window has later been proposed in \cite{cosine_implementation}, also showing that it has satisfactory performance in terms of the error probability.

While existing treatises focus on single-sample-based estimation, in this treatise, we consider the spectral leakage mitigation attained by multiple samples (i.e. measurement outcomes). This is motivated by the fact that the coherence time of \ac{nisq} computers is limited \cite{nisq}, hence classical computing power may be harnessed, by appropriately fusing multiple samples. Explicitly, our contributions are as follows:
\begin{itemize}
\item We derive the \ac{crb} of the quantum phase estimation problem, which sheds light upon the asymptotic multiple-sample performance of practical estimators.
\item We propose a dual-frequency estimator that asymptotically attains the \ac{crb} upon increasing the number of samples. We term this as the asymptotic regime.
\item Using numerical simulations, we demonstrate that the proposed dual-frequency estimator outperform the cosine window-based solution, when the number of samples is sufficiently large.
\end{itemize}

The rest of this treatise is organized as follows. We first formulate the spectral leakage problem and discuss the existing countermeasures in Section \ref{sec:formulation}. Next, we derive the \ac{crb} and propose the dual-frequency estimator in Section \ref{sec:estimator}. The numerical results are then presented in Section \ref{sec:numerical}, and we conclude in Section \ref{sec:conclusion}.

\section{Problem Formulation and Related Works}\label{sec:formulation}
Let us commence by deriving of the output state of the quantum circuit used for phase estimation and shown in Fig.~\ref{fig:qpe_diagram}. After the Hadamard gates, the joint quantum state of both the control register and the target register becomes
\begin{equation}
\ket{+}^{\otimes M} \ket{\psi} = \frac{1}{\sqrt{2^M}} \sum_{n=0}^{2^M-1}\ket{n}\ket{\psi},
\end{equation}
which will be referred to as the \textit{initial state} of the quantum phase estimation algorithm in the rest of this chapter. The subsequent controlled-$\Sop{U}$ gates transform the initial state into the following state
\begin{equation}
\begin{aligned}
\ket{\phi}&=\frac{1}{\sqrt{2^M}}\sum_{n=0}^{2^M-1}\ket{n}\Sop{U}^n\ket{\psi} \\
&=\frac{1}{\sqrt{2^M}}\bigotimes_{k=0}^{M-1} \left(\ket{0}+e^{j 2^k \varphi}\ket{1}\right)\ket{\psi}\\
&=\frac{1}{\sqrt{2^M}} \sum_{n=0}^{2^M-1}e^{jn\varphi}\ket{n}\ket{\psi}.
\end{aligned}
\end{equation}
Note that the inverse quantum Fourier transform $\mathrm{QFT}_M^{-1}$ may be expressed as follows\footnote{In classical signal processing theory, the transform having the minus sign ``$-$'' in the exponents is typically referred to as the discrete Fourier transform, and the transform having the plus sign is referred to as the inverse discrete Fourier transform. However, the quantum computing community is using a different convention, which refers the transform \eqref{iqft} as the inverse quantum Fourier transform \cite[Section 5.1]{ncbook}.}
\begin{equation}\label{iqft}
\mathrm{QFT}_M^{-1}=\frac{1}{\sqrt{N}} \sum_{m=0}^{N-1}e^{-j\frac{2\pi m}{N}}\ket{m}\bra{n},
\end{equation}
where $N=2^M$ is the record length. Hence the quantum state of the control register at the output of the $\mathrm{QFT}_M^{-1}$ may be expressed as
\begin{equation}\label{output}
\ket{\phi}_{\rm out}=\frac{1}{N}\sum_{m=0}^{N-1}\sum_{n=0}^{N-1}e^{jn\left(\varphi-\frac{2\pi m}{N}\right)}\ket{m}.
\end{equation}
The probability of observing the outcome $\ket{y}$ is thus given by
\begin{equation}\label{likelihood_raw}
\begin{aligned}
f(y;\varphi) &= |\bra{y}\phi\rangle_{\rm out}|^2 \\
&= \frac{1}{N^2}\left|\sum_{n=0}^{N-1}e^{jn\left(\varphi-\frac{2\pi y}{N}\right)}\right|^2.
\end{aligned}
\end{equation}
When $\varphi=2\pi k /N,~k\in\mathbb{Z}$, we see that the probability $f(y;\varphi)$ takes the maximum value of $1$ at $y=k$, hence the phase estimation yields the exact solution. However, when $\varphi$ is not an integer multiple of $2\pi/N$, $f(y;\varphi)$ is non-zero for almost all $0\le y\le N-1$, causing large estimation errors. This phenomenon is referred to as ``spectral leakage'' in the literature of classical signal processing \cite{spectral_leakage,spectral_leakage_spl}.

The spectral leakage issue can be mitigated using the classic windowing method \cite{window2}, multiplying the input signal by a ``window'' function defining a weighting vector $\V{\alpha}$. This results in the following observation probability distribution
\begin{equation}\label{likelihood_window}
f(y;\varphi) = \frac{1}{N} \left|\sum_{n=0}^{N-1} \alpha_n e^{jn\left(\varphi-\frac{2\pi y}{N}\right)}\right|^2,
\end{equation}
where $\alpha_n$ is the $n$-th entry of $\V{\alpha}$, and $\|\V{\alpha}\|=1$. Upon comparing \eqref{likelihood_raw} to \eqref{likelihood_window}, we see that the original quantum phase estimation corresponds to a rectangular window of $\V{\alpha}=\frac{1}{\sqrt{N}}\V{1}$. This weighting procedure may be implemented upon replacing the Hadamard gates in the dashed box of Fig.~\ref{fig:qpe_diagram} by customized state preparation circuits \cite{cosine_implementation}.

\begin{figure}[t]
\centering
\includegraphics[width=.485\textwidth]{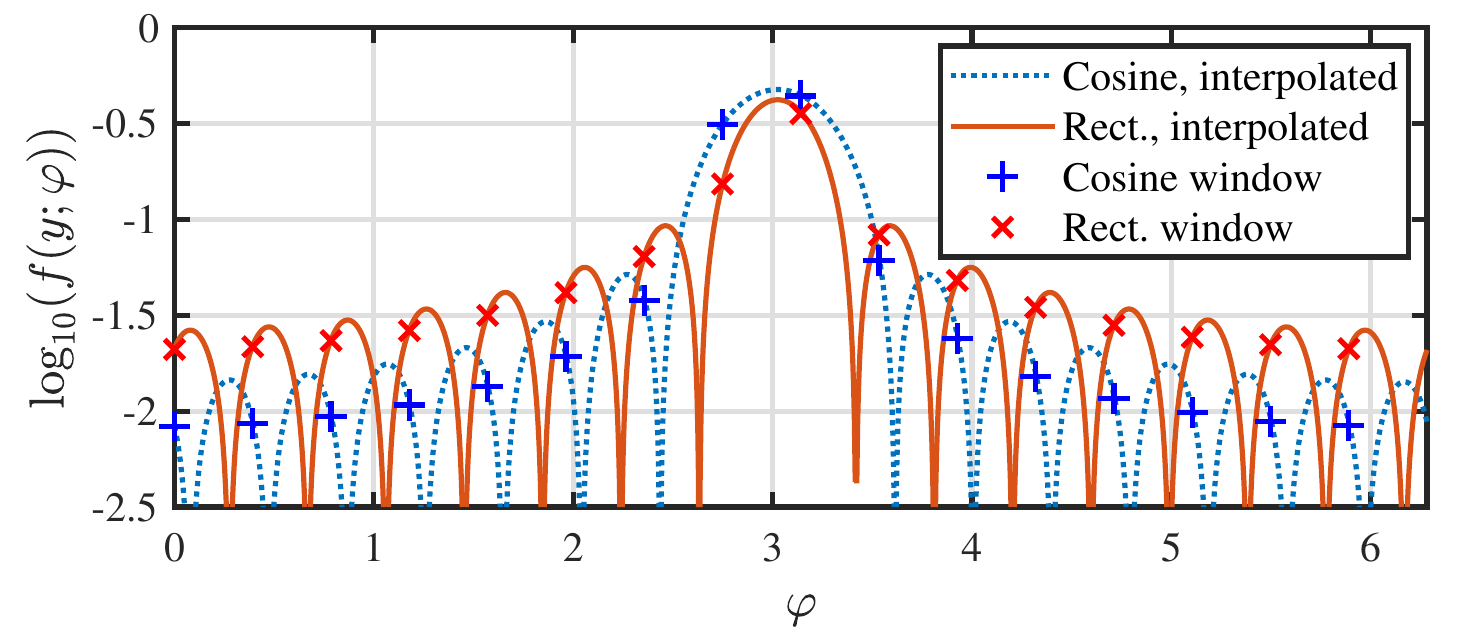}
\caption{Illustration of the observation probability distribution $f(y;\varphi)$ for rectangular and cosine windows, respectively.}
\label{fig:windows}
\end{figure}

It has been shown that in terms of the lowest \ac{mse}, the optimal window is the cosine window \cite{quantum_cosine,classical_cosine}, given by
\begin{equation}
\alpha_n^{(\cos)} = \sqrt{\frac{2}{N}}\sin\left[\frac{\pi (n-1)}{N}\right].
\end{equation}
The authors of \cite{cosine_implementation} have designed an efficient state preparation circuit for the cosine window. The effect of the window may be intuitively interpreted by considering the corresponding observation probability distribution, as portrayed in Fig.~\ref{fig:windows}. Observe that the ``sidelobes'' of the cosine window are much lower than those of the rectangular window, hence the probability that extremely large errors occur is substantially reduced. However, it is also seen that the cosine window has a wider mainlobe. If the sidelobes can be suppressed without widening the mainlobe, we may achieve better performance than that of the cosine window. This motivates our dual-frequency estimator, which will be detailed in Section \ref{sec:estimator}.

When multiple samples are available, it is possible to further improve the accuracy by simply taking the average of these samples, which we refer to as the ``sample-mean estimator''. The sample-mean estimator has a time complexity on the order of $O(N_{\rm s})$, where $N_{\rm s}$ is the number of samples. In fact, when the sidelobes are effectively suppressed, the sample-mean estimator is sufficient for obtaining a near-optimal performance (in the sense of CRB), as will be detailed in Section \ref{sec:estimator}.

\section{Proposed Estimator}\label{sec:estimator}
In this section, we first derive the \ac{crb} of the quantum phase estimation problem, which will be used as a performance benchmark in the numerical simulations of Section \ref{sec:numerical}. Moreover, it also motivates us to propose the dual-frequency estimator detailed in Section \ref{ssec:dual_freq}, which asymptotically approaches the \ac{crb} upon increasing the number of samples. The generic \ac{crb} of the quantum phase estimation problem has been derived in the literature \cite{optimal_qpe,bayesian_qpe}. In this section, we take into account the effect of the window function on the phase estimation performance.

\subsection{\ac{crb} Analysis}\label{ssec:crb}
The \ac{crb} is a lower bound on the mean-squared error of estimators \cite{CRB}, which is tight, when the noise is weak or the number of observations is large. In light of this, it may be used as the metric for determining the optimal window function for a large number of samples.

In general, given the likelihood function $f(y;\theta)$ of an observation $y$, the \ac{crb} of the parameter $\theta$ is given by
\begin{equation}
\mathbb{E}\{(\theta-\hat{\theta})^2\}\ge \frac{1}{\mathrm{FI}(\theta)},
\end{equation}
where
\begin{equation}\label{fisher_information}
\mathrm{FI}(\theta)=\mathbb{E}\left\{\left[\frac{\partial \ln f(y;\theta)}{\partial \theta}\right]^2\right\}.
\end{equation}
is the Fisher information of $\theta$ \cite{CRB}. When there are multiple independent and identical distributed observations, the total Fisher information is the sum of the Fisher information of all individual observations.

In the context of quantum phase estimation, the likelihood function of a single observation is given by \eqref{likelihood_window}, which may be rewritten in a more compact form as
\begin{equation}
\begin{aligned}
f(y;\varphi) &= \frac{1}{N}|\V{e}_y^{\rm H}\V{\alpha}|^2 \\
&= \frac{1}{N}\V{e}_y^{\rm H}\V{\alpha}\V{\alpha}^{\rm H}\V{e}_y,
\end{aligned}
\end{equation}
where
$$
\V{e}_y:=\left[1,~e^{j\left(\varphi-\frac{2\pi y}{N}\right)},\dotsc,~e^{j(N-1)\left(\varphi-\frac{2\pi y}{N}\right)}\right]^{\rm T}
$$
denotes the vector containing all the phase terms in \eqref{likelihood_raw}. For the simplicity of further derivation, we rewrite the Fisher information in \eqref{fisher_information} in the following alternative form
\begin{equation}
\begin{aligned}
\mathrm{FI}(\varphi)&=\mathbb{E}\left\{\left[\frac{\partial \ln f(y;\theta)}{\partial \theta}\right]^2\right\}\\
&=\sum_y f(y;\theta) \left[\frac{\partial \ln f(y;\theta)}{\partial \theta}\right]^2 \\
&=\sum_y \frac{1}{f(y;\varphi)}\cdot \left(\frac{\partial}{\partial \varphi}f(y;\varphi)\right)^2,
\end{aligned}
\end{equation}
where the partial derivative may be expressed as
\begin{equation}
\begin{aligned}
\frac{\partial}{\partial \varphi}f(y;\varphi)&=\frac{1}{N}\left[\left(\frac{\partial \V{e}_y}{\partial \varphi}\right)^{\rm T}\V{\alpha}^*\V{\alpha}^{\rm T}\V{e}_y^*+\left(\frac{\partial \V{e}_y^*}{\partial \varphi}\right)^{\rm T}\V{\alpha}\V{\alpha}^{\rm H}\V{e}_y\right]\\
&=\frac{j}{N}(\V{e}^T\V{N}\V{\alpha}^*\V{\alpha}^T\V{e}^*-\V{e}^H\V{N}\V{\alpha}\V{\alpha}^H\V{e})\\
&=\frac{j}{N}\V{e}_y^{\rm H}(\V{\alpha}\V{\alpha}^{\rm H}\V{N}-\V{N}\V{\alpha}\V{\alpha}^{\rm H})\V{e}_y,
\end{aligned}
\end{equation}
where we have
$$
\M{N}=\mathrm{diag}\{0,1,\dotsc,N-1\}.
$$
Upon introducing $\M{A}=\V{\alpha}\V{\alpha}^{\rm H}$, the Fisher information may be expressed as
\begin{equation}\label{fisher_info}
\begin{aligned}
\mathrm{FI}(\varphi)&=\frac{1}{N}\sum_{y=0}^{N-1}\frac{|\V{e}_y^{\rm H}(\V{\alpha}\V{\alpha}^{\rm H}\V{N}-\V{N}\V{\alpha}\V{\alpha}^{\rm H})\V{e}_y|^2}{\V{e}_y^{\rm H}\V{\alpha}\V{\alpha}^{\rm H}\V{e}_y}\\
&=\frac{1}{N}\sum_{y=0}^{N-1}\frac{|\V{e}^H[\V{A},\V{N}]\V{e}|^2}{\V{e}^H\V{A}\V{e}}\\
&=\frac{4}{N}\sum_{y=0}^{N-1}\frac{\mathrm{Im}^2\{\V{e}_y^{\rm H}\V{A}\V{N}\V{e}_y\}}{\V{e}_y^{\rm H}\V{A}\V{e}_y}.
\end{aligned}
\end{equation}
When there are $N_{\rm s}$ samples, due to their mutual independence, the total Fisher information is simply formulated as $N_{\rm s}\cdot \mathrm{FI}(\varphi)$. This implies that the optimal window in the asymptotic regime of infinite samples may actually be determined by considering the Fisher information of the window applied to a single sample.

\begin{figure}[t]
\centering
\includegraphics[width=.46\textwidth]{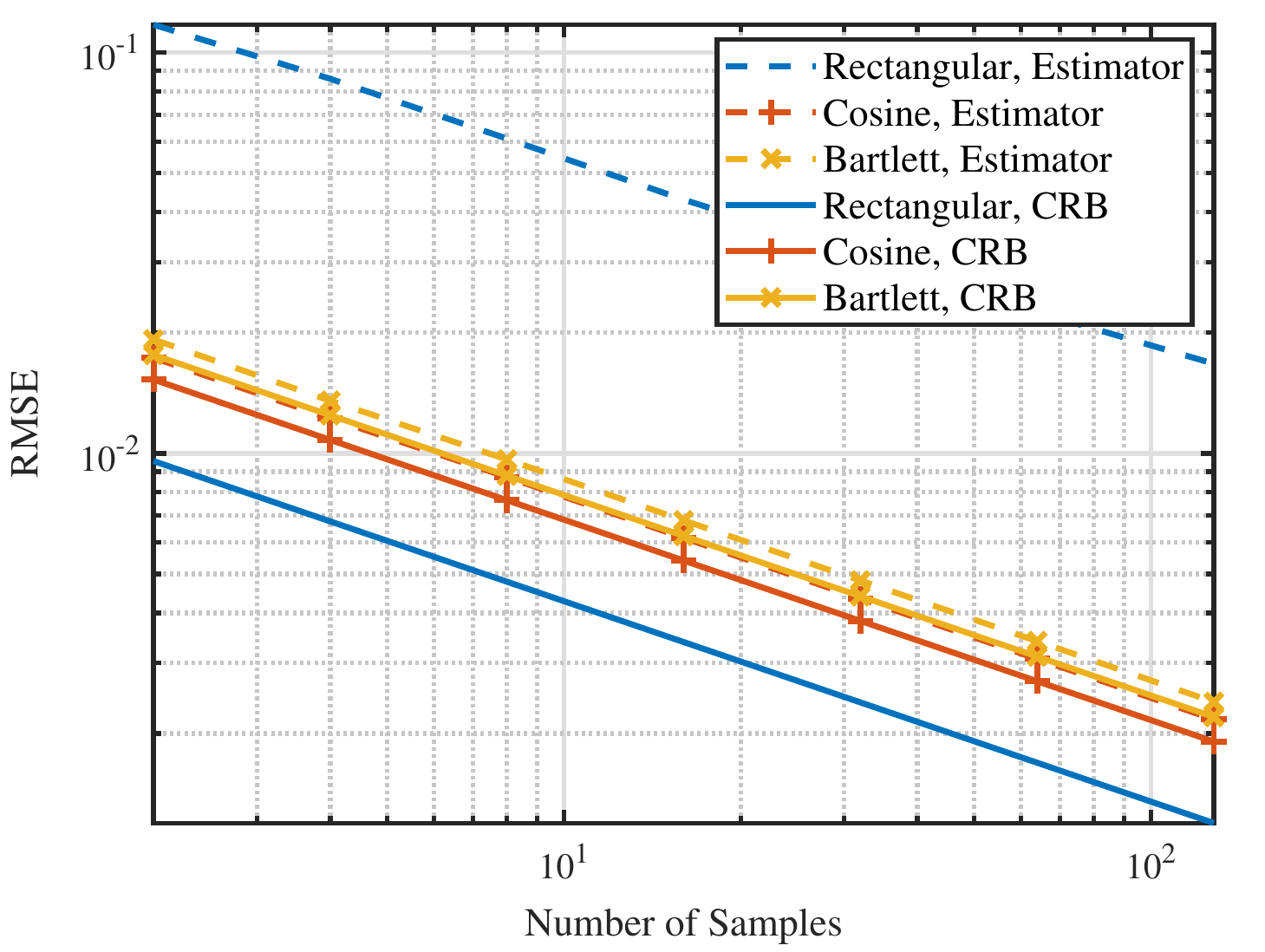}
\caption{The square-root \ac{crb} vs. number of samples of various window functions, compared with the RMSE of corresponding sample-mean estimators.}
\label{fig:samples_crb_vs_algo}
\end{figure}

We may use \eqref{fisher_info} to obtain an intuition about the optimal window in the asymptotic regime. In Fig. \ref{fig:samples_crb_vs_algo}, we compare the square-root \ac{crb} of various windows, where we set the record length to $N=128$, corresponding to $7$ control qubits. Observe that the rectangular window (equivalent to no windowing at all over a finite interval) corresponds to the lowest \ac{crb}. This suggests that it should yield the best asymptotic performance among all window functions. However, we see that the naive sample-mean-based estimator operates far from the \ac{crb}. By contrast, the sample-mean estimators of cosine and Bartlett windows exhibit near-\ac{crb} performances. We will further address this issue in Section \ref{ssec:dual_freq}.

\subsection{The Dual-Frequency Estimator}\label{ssec:dual_freq}
Let us commence our discussion from constructing an approximate maximum likelihood estimator based on the rectangular window with the aid of $N_{\rm s}$ samples. We first denote the samples as $\V{y} = [y_1,~y_2,\dotsc,~y_{N_{\rm s}}]^{\rm T}$. From \eqref{likelihood_raw}, the exact maximum likelihood estimator is given by
\begin{equation}\label{ml_first}
\hat{\varphi}_{\rm ML} = \mathop{\rm argmax}_{\varphi\in[0,2\pi)} ~\sum_{i=1}^{N_s}\ln \left|\sum_{n=0}^{N-1}e^{jn\left(\varphi-\frac{2\pi y_i}{N}\right)}\right|^2.
\end{equation}
Since the samples only take integer values between $0$ and $N-1$, we may use an alternative parametrization of $\V{z}\in\mathbb{R}^{N}$, whose $n$-th entry $z_k$ represents the number of times that $n-1$ occurs in the samples $\V{y}$. Thus we may rewrite \eqref{ml_first} as
\begin{equation}\label{ml_shots}
\hat{\varphi}_{\rm ML} = \mathop{\rm argmax}_{\varphi\in[0,2\pi)} ~\sum_{k=1}^{N}z_k \ln \left|\sum_{n=0}^{N-1}e^{jn\left(\varphi-\frac{2\pi k}{N}\right)}\right|^2.
\end{equation}
A naive strategy of solving \eqref{ml_shots} is an exhaustive search of $\varphi$ in $[0,2\pi)$, which is computationally expansive. To simplify the problem, we first obtain a rough estimate of $\varphi$ as \begin{equation}
\hat{\varphi}_{\rm rough} = \frac{2\pi}{N}\left[(\mathop{\mathrm{argmax}}_k z_k)-1\right].
\end{equation}
Note that the worst-case complexity of this step is on the order of $O(N_{\rm s})$ when $N_{\rm s}\ll N$. Next we conduct a refined search within $[\hat{\varphi}_{\rm rough}-2\pi/N,\hat{\varphi}_{\rm rough}+2\pi/N]$. Although more sophisticated optimization techniques may have better performance, here we consider the simple approach of a uniform grid search over $O(\sqrt{N_{\rm s}})$ grid points, inspired by the fact that the Fisher information is on the order of $O(N_{\rm s})$. To avoid the summation over $n$ in \eqref{ml_shots}, we consider the following approximation for large $N$:
\begin{equation}
\frac{1}{N}\left|\sum_{n=0}^{N-1}e^{jn\left(\varphi-\frac{2\pi k}{N}\right)}\right|\approx \left|\mathrm{sinc}\left(\frac{N\varphi}{2\pi}-k\right)\right|.
\end{equation}
Using the above approximation, the worst-case complexity of the grid search is reduced to $O(N_{\rm s}^{1.5})$. We denote the final estimate as $\hat{\varphi}_{\rm AML}$, which may be obtained as
$$
\hat{\varphi}_{\rm AML} = \mathop{\mathrm{argmax}}_{\substack{\varphi\in\Set{F}\\ \varphi=\hat{\varphi}_{\rm rough}+\frac{4k\pi}{NN_{\rm g}},~k\in\mathbb{Z}}} \sum_{k=1}^N z_k \ln \left|\mathrm{sinc}\left(\frac{N\varphi}{2\pi}-k\right)\right|,
$$
where $N_{\rm g}$ is the number of grid points. The correction term is denoted as $e_{\rm AML} = \hat{\varphi}_{\rm AML}-\hat{\varphi}_{\rm rough}$.

\begin{figure}[t]
\centering
\includegraphics[width=.485\textwidth]{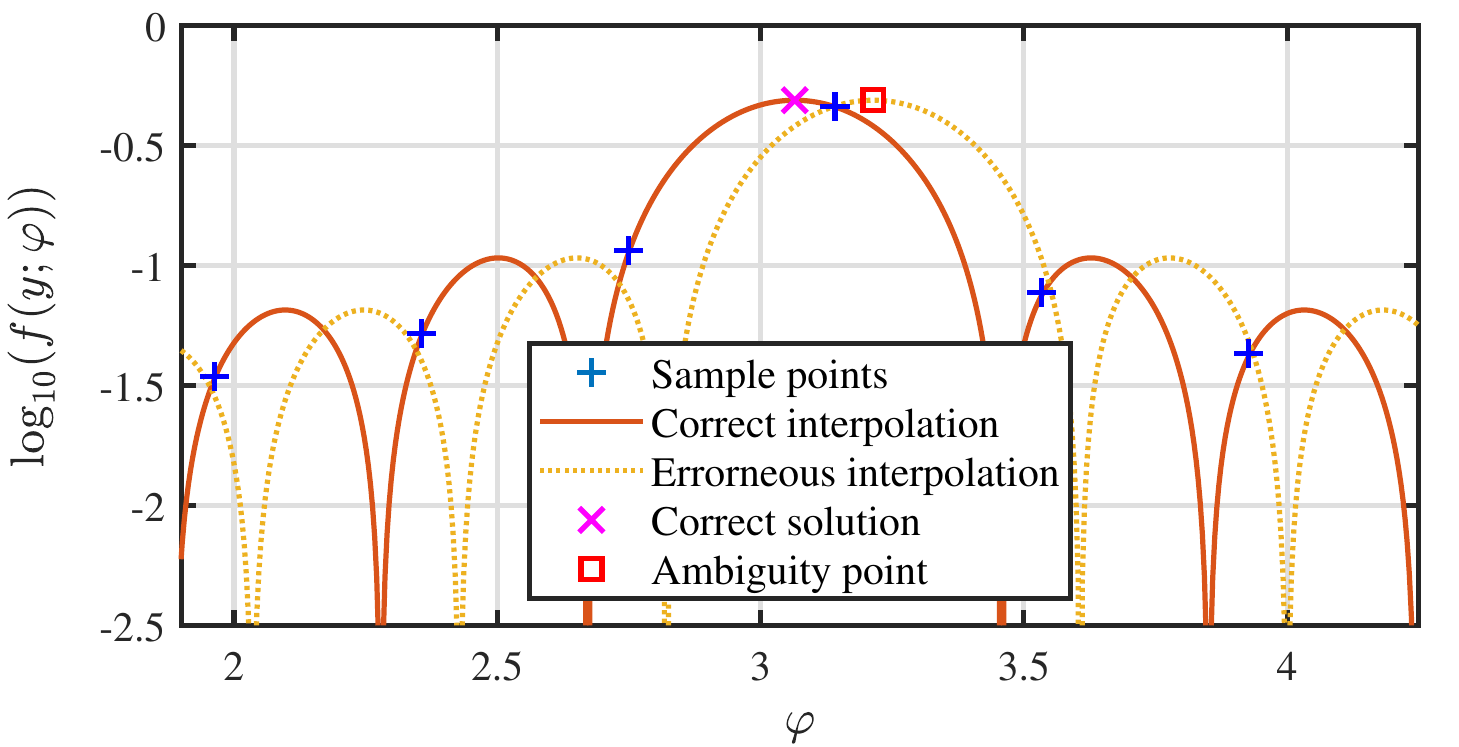}
\caption{Illustration of the ambiguity problem associated with the approximate maximum likelihood estimator when $\varphi$ is close to $2\pi k/N,~k\in\mathbb{Z}$.}
\label{fig:illustration_ambiguity}
\end{figure}

This estimator, however, does not in general yield satisfactory performance. We may develop some further intuition concerning this issue by revisiting Fig.~\ref{fig:windows}. Observe that most of the information about $\varphi$ is contained in the pair of sample points within the mainlobe. When $\varphi$ is near $(2\pi k+1)/N,~k\in\mathbb{Z}$, the largest and the second largest entries in $\V{z}$ may be used as reliable estimates of the two sample points. However, when $\varphi$ is close to $2\pi k/N,~k\in\mathbb{Z}$, it is likely that the second largest entry in $\V{z}$ corresponds to the first sidelobe, causing an unexpected estimation error. As portrayed in Fig.~\ref{fig:illustration_ambiguity}, the difficulty of distinguishing the highest sidelobe from the sample point in the mainlobe having smaller likelihood incurs an ambiguity problem for the maximum likelihood estimator.

\begin{figure}[t]
\centering
\includegraphics[width=.485\textwidth]{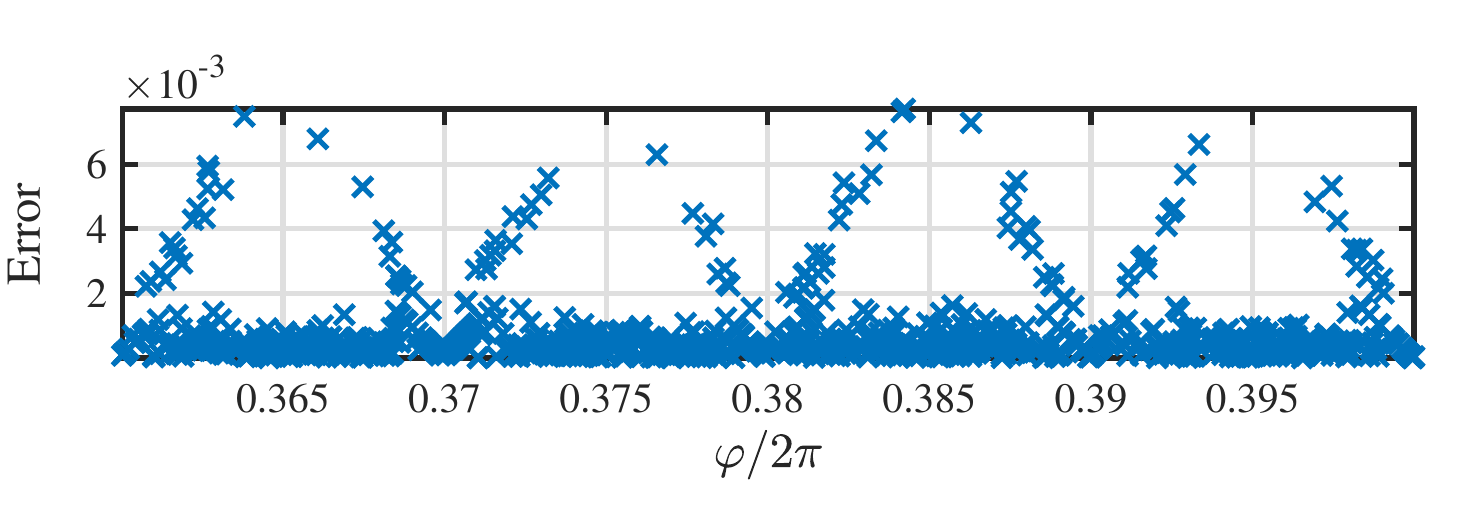}
\caption{Scatter plot of the errors of the approximate maximum likelihood estimator for $\varphi$ uniformly distributed over a fixed interval.}
\label{fig:aml_scatter}
\end{figure}

To understand this ambiguity problem in a more precise manner, let us consider a concrete example. Specifically, we set the record length to $N=100$, the number of samples to $N_{\rm s}=30$, and then plot the corresponding estimation error based on the approximate maximum likelihood estimator characterized in Fig.~\ref{fig:aml_scatter}.\footnote{In practice, the record length can only be an integer power of 2. Here we choose $N=100$ only for the purpose of a clearer illustration.} Observe that the errors are sometimes large when $\varphi$ is close to an integer multiple of $1/2\pi N$. In particular, the magnitude of the error grows linearly with the distance to the closest integer multiple of $1/2\pi N$. As we have discussed previously, this phenomenon originates from the fact that two possible interpolations of the sample points are hardly distinguishable, hence there is a non-zero probability that the estimator yields the erroneous result corresponding to the ``mirror point'' of the correct one across the line of $\varphi = k/2\pi N$, where $k$ is the closest integer multiple of $1/2\pi N$.

To tackle this problem, we split the samples into two sets, each having $N_{\rm s}/2$ samples. For the first set, we compute the aforementioned approximate maximum likelihood estimator, and obtain $\hat{\varphi}_{{\rm rough},1}$ and $e_{{\rm AML},1}$. For the second set, we apply a frequency offset of $1/2N$ (one half of a frequency resolution unit) to the control register, resulting in the following maximum likelihood problem
\begin{equation}\label{ml_shots2}
\begin{aligned}
\hat{\varphi}_{{\rm ML},2} &= \mathop{\rm argmax}_{\varphi\in[0,2\pi)} ~\sum_{k=1}^{N}z_k \ln \left|\sum_{n=0}^{N-1}e^{jn\left(\varphi-\frac{2\pi k}{N}+\frac{\pi}{N}\right)}\right|^2\\
&\hspace{3mm}+\frac{\pi}{N}.
\end{aligned}
\end{equation}
Similarly, we may obtain $\hat{\varphi}_{{\rm rough},2}$ and $e_{{\rm AML},2}$. Based on these results, we construct four candidate intermediate estimates, contained in a vector as follows:
\begin{equation}\label{candidate_estimates}
\begin{aligned}
\V{u} &\!=\! \Big[\hat{\varphi}_{{\rm rough},1}\!+\!e_{{\rm AML},1},~\hat{\varphi}_{{\rm rough},1}\!-\!e_{{\rm AML},1}, \\
&\hspace{7mm}\!\hat{\varphi}_{{\rm rough},2}\!+\!e_{{\rm AML},2},~\hat{\varphi}_{{\rm rough},2}\!-\!e_{{\rm AML},2}+\frac{2\pi}{N}\Big]^{\rm T}.
\end{aligned}
\end{equation}
Finally, we find the pair of entries in $\V{u}$ that are the closest to each other, meaning that
\begin{equation}\label{matching}
(i,j) = \mathop{\mathrm{argmax}}_{(i,j)} ~(u_i-u_j)^2,
\end{equation}
and the final estimate is given by
\begin{equation}
\hat{\varphi}_{\rm DF} = \frac{1}{2}(u_i+u_j).
\end{equation}
The overall complexity of the dual-frequency estimator is at most $O(N_{\rm s}^{1.5})$, which is comparable to the $O(N_{\rm s})$ complexity of the simple sample-mean estimator. In practice, the complexity can be lower than $O(N_{\rm s}^{1.5})$. To elaborate, note that the actual complexity of the dual-frequency estimator is proportional to
\begin{equation}
C_{\rm DF} = N_{\rm bin}\sqrt{N_{\rm s}},
\end{equation}
where $N_{\rm bin}$ denotes the number of non-empty ``bins'' $z_k\neq 0$. In fact, most of the bins would be empty, since the probability of observing those outcomes would be extremely low. Moreover, the bins containing a small number of samples would provide less information about the desired phase $\varphi$, since they are much noisier than the bins containing more samples. For the numerical simulations presented in this chapter, we choose $8$ bins having the largest number of samples. In light of this, the actual complexity of the dual-frequency estimator becomes $O(N_{\rm s}+\sqrt{N_{\rm }})$, where the $O(N_{\rm s})$ comes from the fact that re-organizing the samples into bins would require $O(N_{\rm s})$ operations.

\begin{figure}[t]
\centering
\includegraphics[width=.485\textwidth]{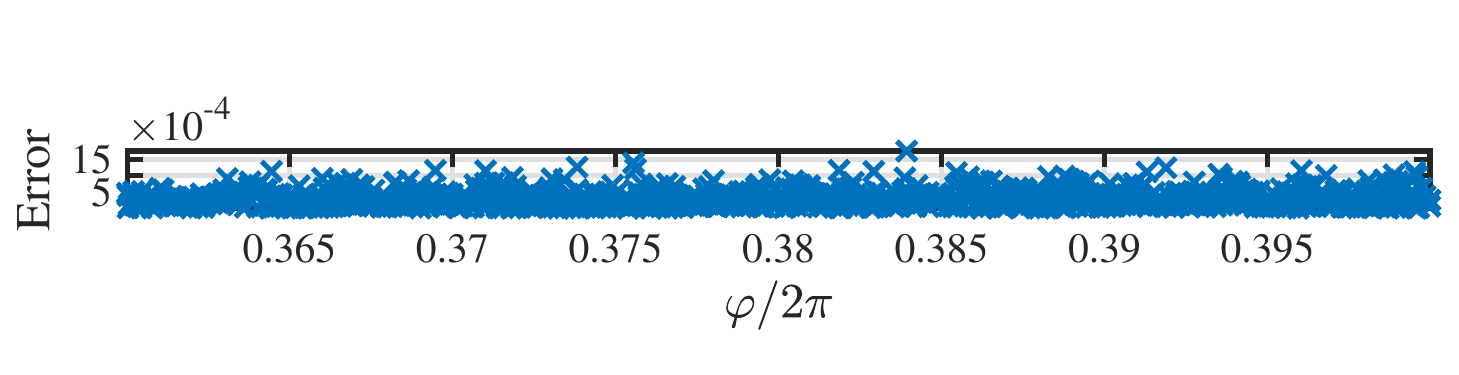}
\caption{Scatter plot of the errors of the dual-frequency estimator for $\varphi$ uniformly distributed over a fixed interval.}
\label{fig:df_scatter}
\end{figure}

Next we show that the dual-frequency estimator indeed resolves the ambiguity problem using a concrete example. Similar to Fig.~\ref{fig:aml_scatter}, we set the record length to $N=100$ and the number of samples to $N_{\rm s}=30$. The estimation error of the dual-frequency estimator is portrayed in Fig.~\ref{fig:df_scatter}. As seen from the figure, the dual-frequency estimator only produces small errors, and the linearly increasing trend of errors seen in Fig.~\ref{fig:aml_scatter} does not appear. This implies that the ambiguity point becomes distinguishable from the correct solution.

The rationale of the dual-frequency estimator may be intuitively understood as follows. To circumvent the difficulty of distinguishing the correct solution from the ambiguity point, we include both points into the set of candidate estimates, and also apply the same technique to the set of samples associated with the $(1/2N)$-frequency offset. Since the ambiguity points of both sets are unlikely to be the same due to the frequency offset, we may then identify the correct solution by finding the matching pair of candidate solutions using \eqref{matching}.

Finally, we show that the $(1/2N)$-frequency offset exploited in the estimator may be implemented using single-qubit phase rotation gates, hence the computational overhead of state preparation is negligible. Note that this offset may be obtained by initializing the input state of the control register as
\begin{equation}
\ket{\phi}_{\rm in} = \frac{1}{\sqrt{N}}\sum_{n=0}^{N-1} e^{\frac{j\pi n}{N}}\ket{n}.
\end{equation}
This state admits the following simple tensor-product form
\begin{equation}
\ket{\phi}_{\rm in} = \bigotimes_{m=1}^{M} \frac{1}{\sqrt{2}}\left(\ket{0} + e^{\frac{j\pi 2^{m-1}}{N}}\ket{1}\right),
\end{equation}
which may be implemented by applying the phase rotation gate $\Sop{R}_{\rm z}\left(\pi N^{-1}2^{m-1}\right)$ to each $m$-th qubit in the control register. This method bears some resemblance with the controlled-$\Sop{U}$ gates in the quantum phase estimation circuit, which also constructs a sinusoidal signal (in other words, a frequency shift) on the control register. This implies that the qubit complexity of the dual-frequency estimator is $\lceil \log_2 N\rceil$, which is the same as that of the windowing methods.

\section{Numerical Results}\label{sec:numerical}
\begin{figure}[t]
\centering
\includegraphics[width=.45\textwidth]{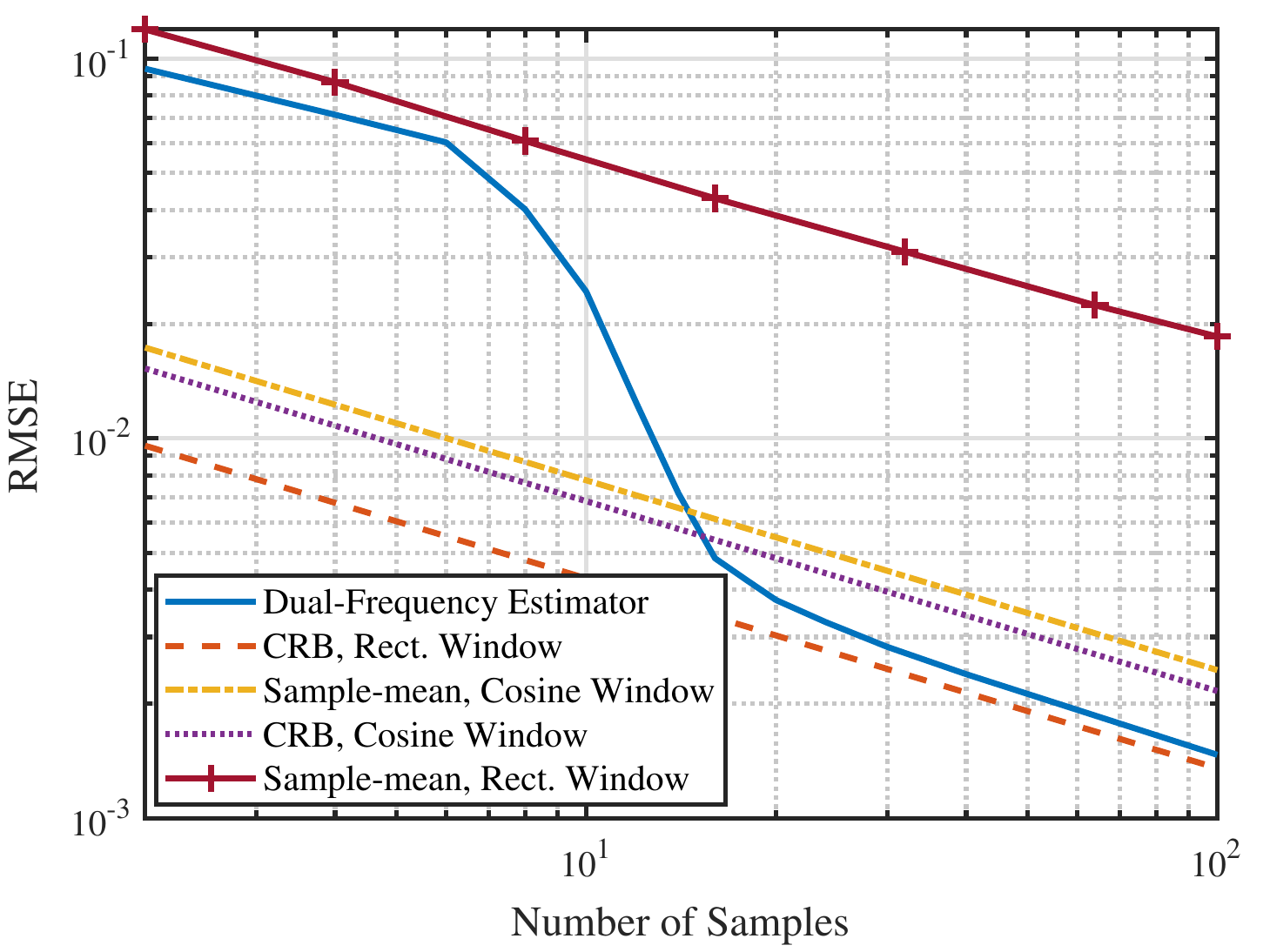}
\caption{The phase estimation RMSE vs. the number of samples of both the proposed dual-frequency estimator and the cosine window method, compared to the corresponding \ac{crb}s.}
\label{fig:crb_vs_algo}
\end{figure}

In this section, we characterize the performance of the proposed estimator using numerical simulations. We first consider the relationship between the \ac{rmse} of the estimators and the number of samples. In this example, we set the number of control qubits to $M=7$, corresponding to $N=128$. The number of samples $N_{\rm s}$ varies from $2$ to $100$. The unknown phase $\varphi$ is randomly chosen from $(0,2\pi)$ in each of the $10^5$ Monte Carlo trails. The \ac{rmse} of the dual-frequency estimator as well as that of the sample-mean estimator based on the cosine window is portrayed in Fig.~\ref{fig:crb_vs_algo}, where the corresponding \ac{crb}s are also incorporated as benchmarks.

Observe that all curves in Fig.~\ref{fig:crb_vs_algo} exhibit the same linear trend in the asymptotic regime on the log-log scale. This implies that both estimators have the same asymptotic error scaling as the \ac{crb}, which scales on the order of $O(1/\sqrt{N_{\rm s}})$ (in terms of \ac{rmse}). When $N_{\rm s}$ is insufficiently large, the dual-frequency estimator exhibits a performance similar to that of the sample-mean estimator. But as $N_{\rm s}$ increases, the performance of the dual-frequency estimator ``switches'' to near-\ac{crb} operation. Observe furthermore that it outperforms the cosine window-based method for $N_{\rm s}\ge 16$. The relatively low accuracy of the dual-frequency estimator in the small-$N_{\rm s}$ regime originates from the fact that the number of samples is to small for us to construct any reliable candidate estimate (as given in \eqref{candidate_estimates}). Consequently, the sample points that are far away from the true value of $\varphi$ (i.e., the ``outliers'') cannot be reliably identified, hence would cause large estimation errors.

\begin{figure}[t]
\centering
\includegraphics[width=.45\textwidth]{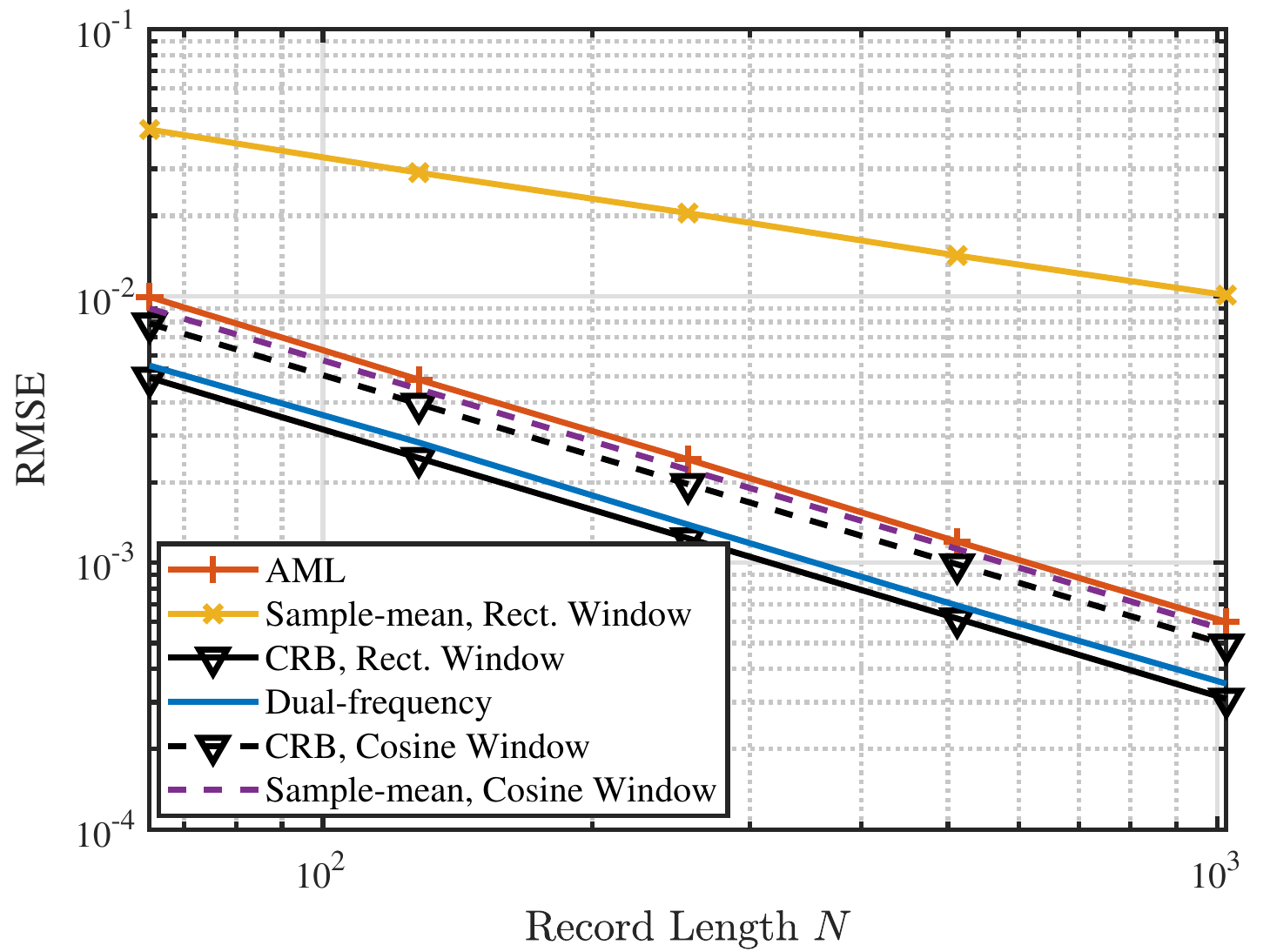}
\caption{The phase estimation RMSE vs. the record length $N$ of the proposed dual-frequency estimator, the approximate maximum likelihood estimator, and the cosine window method, compared to the corresponding \ac{crb}s.}
\label{fig:crb_vs_algo_vs_N}
\end{figure}

Next we consider the dependence of the \ac{rmse} on the record length $N$. We set the number of samples to $N_{\rm s}=30$. The number of control qubits varies from $6$ to $10$, corresponding to $N=64, 128, 256, 512, 1024$. The corresponding results are portrayed in Fig.~\ref{fig:crb_vs_algo_vs_N}. We also incorporate the approximate maximum likelihood estimator in Fig.~\ref{fig:crb_vs_algo_vs_N} for a better illustration.

Observe that the sample-mean estimator based on the rectangular window exhibits an $O(1/\sqrt{N})$ scaling, while the others exhibit $O(1/N)$ scaling. This is a phenomenon that has also been observed in \cite{quantum_cosine}, which suggests that the rectangular window does not provide a substantial quantum speedup in the sense of \ac{rmse} scaling, since the $O(1/N)$ scaling (i.e., the ``Heisenberg limit \cite{hlimit}) is an important characteristic of quantum algorithms conceived for phase estimation.

We also observe that the dual-frequency estimator does not exhibit the aforementioned bimodal phenomenon with respect to $N$. This suggests that for large $N$, we may still use a constant number of samples (for example, $N_{\rm s} \ge 20$ as indicated by Fig.~\ref{fig:crb_vs_algo}) to achieve a near-\ac{crb} performance.

\begin{figure}[t]
\centering
\includegraphics[width=.45\textwidth]{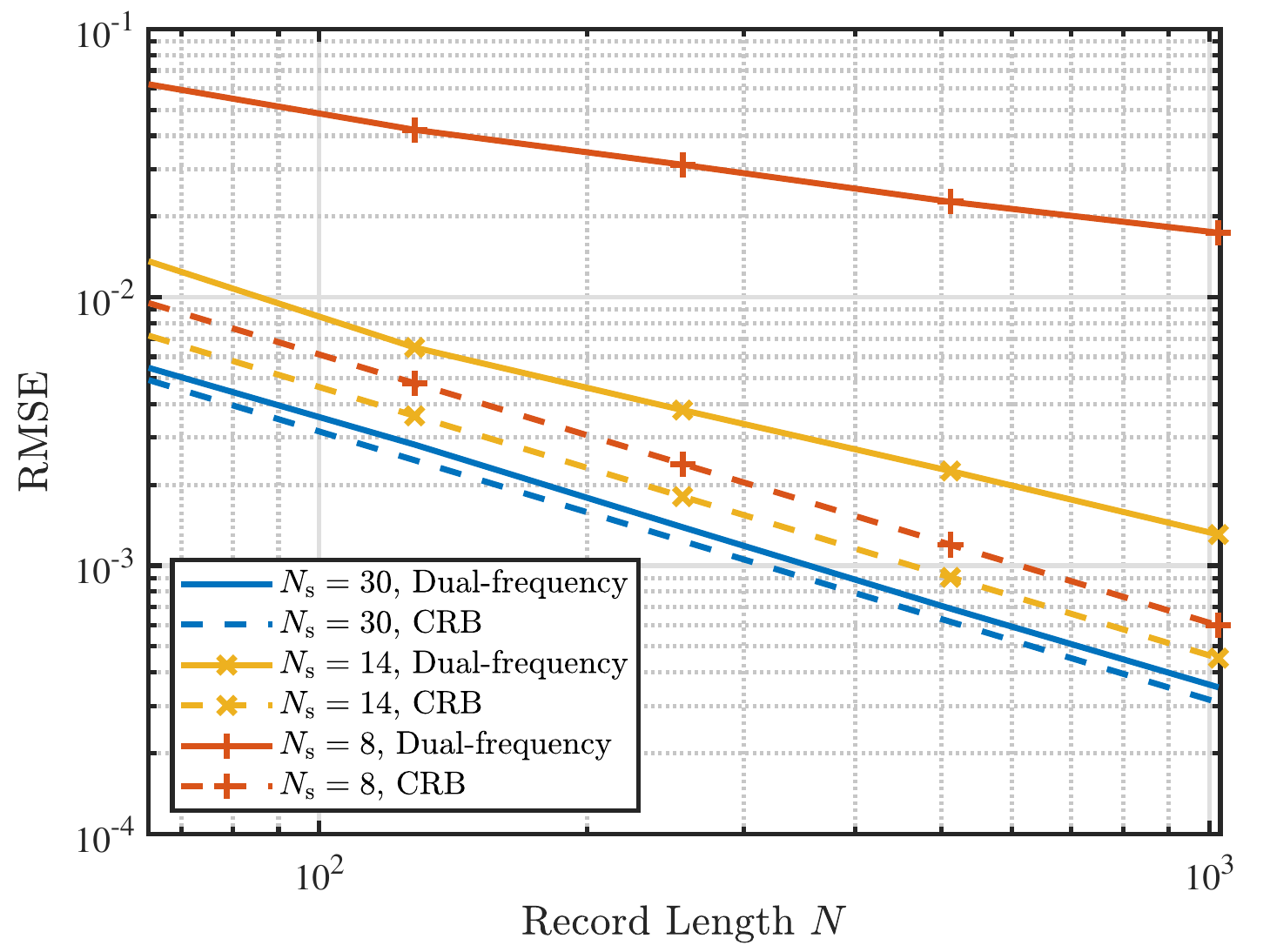}
\caption{The \ac{rmse} of the dual-frequency estimator vs. the record length, for $N_{\rm s}$ values in different regions.}
\label{fig:df_various_Ns}
\end{figure}

In Fig.~\ref{fig:df_various_Ns}, we portray the dependence of the RMSE on the record length $N$ for different values of $N_{\rm s}$. Specifically, we pick three values of $N_{\rm s}$ from the \textit{ambiguity region} ($N_{\rm s}\le 10)$, the \textit{transition region} ($10<N_{\rm s}\le 20$), and the \textit{asymptotic region} ($N_{\rm s}>20$). We observe that in the asymptotic region ($N_{\rm s}=30$), the dual-frequency estimator exhibits a near-CRB performance, and has an order of $O(1/N)$ RMSE scaling. By contrast, in the ambiguity region, the estimator exhibits an order of  $O(1/\sqrt{N})$ RMSE scaling. Finally, in the transition region, the order of RMSE scaling is between $O(1/\sqrt{N})$ and $O(1/N)$.

\section{Conclusions}\label{sec:conclusion}
A dual-frequency phase estimator was proposed for mitigating the spectral leakage-induced error in the quantum phase estimation algorithm based on multiple samples. We have presented its \ac{crb} analysis in the asymptotic regime. This \ac{crb} also inspires the design of our dual-frequency estimator. Compared to the naive sample-mean estimator, the proposed estimator attains the Heisenberg limit in the sense of exhibiting \ac{rmse} scaling on the order of $O(1/N)$. Furthermore, the estimator is capable of outperforming the cosine window, which is shown to be optimal for single-sample estimation, when the number of samples is sufficiently large (but constant with respect to $N$). Our method may inspire future research on the algorithmic error mitigation for a broader range of quantum algorithms.

\bibliographystyle{ieeetran}
\bibliography{IEEEabrv,QEM}

\end{document}